\documentclass[prl,nofootinbib,twocolumn]{revtex4}
\usepackage{graphicx}
\usepackage{bm}
\usepackage{float}
\usepackage{subfigure}
\usepackage{amsmath}
\usepackage{amsfonts}
\usepackage{epsfig}
\usepackage{hyperref}
\usepackage{mathrsfs}

\def\bed{\begin{description}}
\def\eed{\end{description}}

\def\ba{\begin{array}}
\def\ea{\end{array}}

\def\u1{$U(1)$}
\def\suu1{$SU(2)\times U(1)$}

\def\nn{\nonumber}

\begin{document}
\newcommand{\newc}{\newcommand}
\newc{\be}{\begin{equation}}
\newc{\ee}{\end{equation}}
\newc{\bear}{\begin{eqnarray}}
\newc{\eear}{\end{eqnarray}}
\newc{\bea}{\begin{eqnarray*}}
\newc{\eea}{\end{eqnarray*}}
\newc{\D}{\partial}
\newc{\ie}{{\it i.e.} }
\newc{\eg}{{\it e.g.} }
\newc{\etc}{{\it etc.} }
{\newc{\etal}{{\it et al.}}
\newc{\lcdm}{$\Lambda$CDM}
\newc{\ra}{\rightarrow}
\newc{\lra}{\leftrightarrow}
\newc{\lsim}{\buildrel{<}\over{\sim}}
\newc{\gsim}{\buildrel{>}\over{\sim}}
\newcommand{\fs}{{\rm{\it f\sigma_8}}}
\newcommand{\mincir}{\raise
-3.truept\hbox{\rlap{\hbox{$\sim$}}\raise4.truept\hbox{$<$}\ }}
\newcommand{\magcir}{\raise
-3.truept\hbox{\rlap{\hbox{$\sim$}}\raise4.truept\hbox{$>$}\ }}

\title{Stabilizing the Semilocal String with a Dilatonic Coupling}

\author{Leandros Perivolaropoulos}\email{leandros@uoi.gr}

\author{Nikos Platis}\email{n\_platis@yahoo.com}
\affiliation{Department of Physics, University of Ioannina, Greece}

\date{\today}

\begin{abstract}
We demonstrate that the stability of the semilocal vortex can be significantly improved by the presence of a dilatonic coupling of the form $e^\frac{q \vert \Phi \vert^2}{\eta^2} F_{\mu \nu}F^{\mu \nu}$ with $q>0$ where $\eta$ is the scale of symmetry breaking that gives rise to the vortex. For $q=0$ we obtain the usual embedded (semilocal) Nielsen-Olesen vortex. We find the stability region of the parameter $\beta \equiv (\frac{m_\Phi}{m_A})^2$  ($m_\Phi$ and $m_A$ are the masses of the scalar and gauge fields respectively). We show that the stability region of $\beta$ is $0<\beta<\beta_{max}(q)$ where $\beta_{max}(q=0)=1$ (as expected) and $\beta_{max}(q)$ is an increasing function of $q$. This result may have significant implications for the stability of the electroweak vortex in the presence of a dilatonic coupling (dilatonic electroweak vortex).
\end{abstract}
\maketitle
The Nielsen-Olesen (NO) vortex \cite{NieOle73,Peri93} is a topologically stable static solution of the Abelian-Higgs model. The Lagrangian density of this model is of the form
\be
\mathscr{L}=- \frac{1}{4 e^2} F_{\mu \nu} F^{\mu \nu} + |D_{\mu} \Phi|^2 - V( \vert \Phi \vert ^2)
\label{abhiggslang}
\ee
where $\Phi$ is a complex scalar field, $V(\Phi)= \frac{\lambda}{4} \left(|\Phi|^2 - \eta^2 \right)^2$, $D_{\mu} = \D_{\mu} - i A_{\mu}$ and $F_{\mu \nu}=\D_{\mu}A_{\nu} - \D_{\nu}A_{\mu}$.

The NO vortex ansatz is of the form\footnote{In the numerical analysis of this study we have set the winding number $m=1$.}
\bear
\Phi&=&f(r) e^{im\theta} \label{phianz} \\
A_\theta &=& a(r) \label{athetaanz}
\eear
Variation of the Lagrangian (\ref{abhiggslang}) leads to the field equations for $f(r)$ and $a(r)$ as
\bear
& f'' + \frac{f'}{r} - \frac{f}{r^2}(m - u)^2 - \frac{\lambda}{2 e^2}(f^2 -1)f  =  0 \label{feq} \\
& u'' - \frac{u'}{r} + 2 f^2 (m-u)  =  0 \label{ueq}
\eear
where  $u\equiv a(r) r$ and we have implemented the following rescaling:
\bear
f \to \bar{f} & = & \eta f \label{rescalef} \\
r \to \bar{r} & = & \frac{r}{\eta e} \label{rescaler}
\eear
The NO boundary conditions to be imposed on (\ref{feq}) and (\ref{ueq}) are $f(0)=u(0)=0$, $f(r\rightarrow \infty)=1$ and $u(r\rightarrow \infty) = m$. Clearly, the NO solution for $f(r)$, $u(r)$ depends on a single parameter $\beta \equiv \frac{\lambda}{2 e^2}$ which is the squared ratio of the scalar field mass $m_\Phi = \frac{\sqrt{\lambda} \eta}{\sqrt{2}}$ over the gauge field mass $m_A=e \eta$.

The energy density of the NO vortex is of the form
\be
\rho  =  f'^2 + \frac{f^2}{r^2}(m-u)^2 + \frac{u'^2}{2r^2} + \frac{\beta}{2}(f^2 - 1)^2 \label{noendens}
\ee
The NO vortex solution can also be embedded in generalizations of the Abelian-Higgs model. For example the semilocal Lagrangian \cite{VacAch91}
\be
\mathscr{L}=- \frac{1}{4 e^2} F_{\mu \nu} F^{\mu \nu} + (D_{\mu} \Phi)^\dagger (D_{\mu} \Phi)- V(\Phi^\dagger \Phi)
\label{semiloclang}
\ee
is obtained by promoting the $U(1)_{gauge}$ symmetry of the Abeian-Higgs model to an $SU(2)_{global}\times U(1)_{gauge}$ symmetry. This is achieved by replacing the complex scalr $\Phi$ by a complex doublet
\be
\Phi = \begin{pmatrix} \Phi_1 \cr \Phi_2 \cr\end{pmatrix} \label{phidouplet}
\ee
The embedded NO vortex ansatz (semilocal vortex) is of the form
\be
\Phi = \begin{pmatrix} 0 \cr f(r) e^{im\theta} \cr\end{pmatrix} \label{phidouplet}
\ee
while for the gauge field eq. (\ref{athetaanz}) remains unchanged. By varying the semilocal Lagrangian it is easy to show that the field equations obeyed by $f(r)$ and $a(r)$ (or $u(r) \equiv a(r) r$) are identical to the NO equations (\ref{feq}) and (\ref{ueq}). Thus the NO vortex solution is {\it embedded} in the generalized semilocal Lagrangian. However, due to the $S^3$ topology of the semilocal vacuum, the stability of the embedded vortex is not topological. It is only dynamical and is valid for a finite range of the parameter $\beta$. It may be shown \cite{Hin92,Hin93,AchKuiPerVac92,AchBorLid98} that this range of stability is $0<\beta <1$.

The NO vortex can be embedded in several other generalizations of the Abelian-Higgs model which involve broken U(1) symmetries. For example it can be embedded in the bosonic sector of standard Glashow-Salam-Weinberg (GSW) electroweak model \cite{GlaSalWei} with $SU(2)_L\times U(1)_Y$ symmetry. One type of such embedded vortices is also known as the {\it electroweak $Z$-vortex}\cite{Nam77,Vachaspati:1992jk,JamPerVac92}. There is a parameter region of dynamical stability of the electroweak $Z$-vortex. It  is determined by two parameters: the squared ratio $\beta$ of the Higgs mass $m_H$ over the $Z_\mu$ mass $m_Z$ ($\beta \equiv (\frac{m_H}{m_Z})^2$) and the Weinberg angle $\theta_w$\cite{JamPerVac92}. Thus, the stability range of the embedded electroweak $Z$-vortex is of the form $0<\beta<\beta_{max}(\theta_w)$. For $\theta_w=\frac{\pi}{2}$ the bosonic sector of the GSW Lagrangian reduces to the semilocal Lagrangian and therefore $\beta_{max}(\theta_w=\frac{\pi}{2})=1$. For $\theta_w < \frac{\pi}{2}$ we have $\beta_{max}(\theta_w)<1$ and therefore the stability range decreases. The experimental values $\beta=\left(\frac{m_H}{m_Z}\right)^2=(\frac{91.2 GeV}{125 GeV})^2 \simeq 0.53$ and $sin\theta_w=0.23$ are outside of the stability region \cite{JamPerVac92}.

There has been a wide range of studies aiming at constructing models where the stability of the semilocal and/or electroweak vortex is improved. These studies have attempted to improve the stability using thermal effects \cite{HolHsuVacWat92}, extra scalar fields \cite{EarJam93,VacWat93}, external magnetic fields \cite{GarMon95}, fermions \cite{EarPer94,Nac95} or spinning scalar field in a charged background \cite{Per94}. Most of these studies have either lead to models where the vortex stability is worse than the usual semilocal Lagrangian or to particularly contrived models requiring external backgrounds. Even for the classical stability region of semilocal and electroweak strings, there are instabilities at the quantum level \cite{PreVil92}.

In this study we attempt to increase the stability region of the semilocal vortex by considering a simple and generic generalization of the Abelian-Higgs model Lagrangian: the {\it dilatonic Abelian-Higgs model} defined to be of the form
\be
\mathscr{L} = \vert D_{\mu} \Phi \vert^2  - \frac{B(\vert \Phi \vert ^2)}{4 e^2}F_{\mu \nu}F^{\mu \nu} - \frac{\lambda}{4}(\vert \Phi \vert ^2 - \eta^2)^2
\label{dilatoniclang}
\ee
where $B(\vert \Phi \vert ^2)= e^\frac{q\vert \Phi \vert ^2}{\eta^2} $ is a dilatonic coupling that allows for dynamics of the effective gauge coupling $\frac{e}{\sqrt{B(\vert \Phi\vert ^2)}}$. In the limit $B(\vert \Phi\vert ^2)\rightarrow 1$ ($q\rightarrow 0$) we obtain the usual Abelian-Higgs model with the NO vortex solution.
\begin{figure}[!ht]
\centering
\includegraphics[scale=0.4]{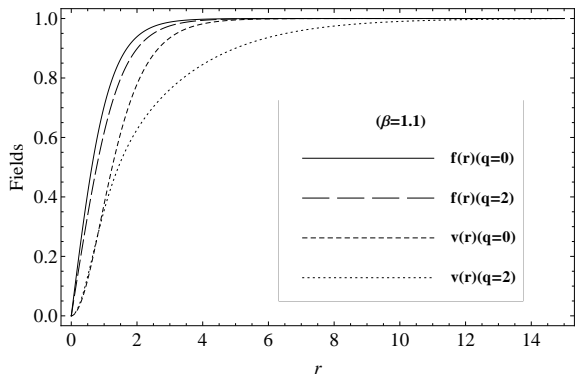}
\caption{Solutions for $f(r)$, $u(r)$ for the dilatonic  gauged vortex for $\beta=1.1$, $q=0$ and $q=2$ respectively. Notice the thickness increase as we increase $q$. It is due to the amplified weight of the gauge field kinetic term in the energy density of the vortex (a reduced gauge field gradient in regions where $f$ is large 'saves' energy).}
\label{fig:dilatonicsols}
\end{figure}
Considering now the NO ansatz in the dilatonic Abelian-Higgs model we obtain the rescaled field equations for the dilatonic gauge vortex
\bear
f'' + \frac{f'}{r} - \frac{f}{r^2}(m-u)^2 - \frac{1}{2} q e^{q f^2}  \left(\frac{u'}{r} \right)^2 f &-& \nonumber \\  -  \beta(f^2  -1)f & = & 0 \label{dilatoncnof}\\
u'' - \frac{u'}{r} + 2  f^2 e^{-qf^2}(m-u) & = & 0 \label{dilatonicnou}
\eear
where as usual $\beta = \frac{\lambda}{2e^2}$.
The corresponding energy density is of the form
\be
\rho =  f'^2 + \frac{f^2}{r^2}(m-u)^2  + \frac{1}{2}e^{q f^2}\left(\frac{u'}{r} \right)^2 + \frac{\beta}{2}(f^2 -1)^2 \label{rhodilatno}
\ee
Using the NO boundary conditions, it is straightforward to obtain the dilatonic vortex solution of equations (\ref{dilatoncnof}), (\ref{dilatonicnou}) for various values of the parameters $q$ and $\beta$ (Fig \ref{fig:dilatonicsols}). A simple mathematica code for the derivation of this solution, based on the minimization of the energy density (\ref{rhodilatno}), is provided in the Appendix.

In order to investigate the stability of the embedded dilatonic vortex we generalize the dilatonic Abelian-Higgs Lagrangian to a dilatonic semilocal Lagrangian with $SU(2)_{global}\times U(1)_{gauge}$ symmetry.
\be
\mathscr{L} =  (D_{\mu} \Phi)^\dagger (D_{\mu} \Phi) - \frac{B(\Phi^\dagger \Phi)}{4 e^2}F_{\mu \nu}F^{\mu \nu} - \frac{\lambda}{4}(\Phi^\dagger \Phi - \eta^2)^2
\label{dilatonicembno}
\ee
We then consider the perturbed fields $\Phi$ and $A_\mu$ as
\be
\Phi = \begin{pmatrix} g \cr f e^{im\theta} + \delta\Phi_2 \cr\end{pmatrix} \label{dilatonicphipert}
\ee
\be
A_\mu = (\delta A_0,\delta A_r, A_\theta+\delta A_\theta, \delta A_z) \label{dilatonapert}
\ee
The energy perturbations due to $\delta \Phi_2$ and $\delta A_\mu$ decouple and can only lead to increase of the embedded dilatonic vortex energy. The corresponding energy perturbation is identical to the energy perturbation of the topologically stable NO vortex and therefore it is positive definite. Thus the stability of the dilatonic embedded vortex is determined by the energy perturbation due to $\delta \Phi_1 \equiv g$.
The energy of the perturbed vortex is of the form
\bear
E_g &=& \int_0^\infty dr \; r\;  ( g'^2 + f'^2 + \frac{f^2}{r^2}(m-u)^2 + \frac{u^2 g^2}{r^2} + \nn \\ &+& \frac{1}{2}e^{q(f^2 +g^2)}\left(\frac{u'}{r} \right)^2 + \frac{\beta}{2}(f^2 + g^2 -1)^2) \nn \\ &\equiv & E_0 + \delta E_g \label{egdilatonic}
\eear
where we have included only perturbations due to $g$ and $E_0$ is the unperturbed energy of the embedded dilatonic vortex.
The energy perturbation due to $g$ may be written in the form
\be
\delta E_g = \int_0^\infty dr \; r \; (g {\hat O} g) \label{degvso}
\ee
where ${\hat O}$ is a Schrodinger-like Hermitian operator of the form
\be
{\hat O}  =  -\frac{1}{r} \frac{d}{dr}\left(r\frac{d}{dr} \right) + \frac{u^2}{r^2} + \frac{q}{2}\left(\frac{u'}{r} \right)^2 +  \beta (f^2 -1)\label{odef}
\ee
and we have only kept terms up to second order in $g$. The Schrodinger potential corresponding to ${\hat O}$ is
\be
V_{\it Schrodinger}  =  \frac{u^2}{r^2} + \frac{q}{2}\left(\frac{u'}{r} \right)^2 +  \beta (f^2 -1) \label{ptlschr}
\ee
For values of the parameters $q$ and $\beta$ for which $\hat O$ has no negative eigenvalues  we have $\delta E_g \geq 0$ and therefore no instability develops. In order to determine if $\hat O$ has negative eigenvalues we may solve ${\hat O} g(r) =0$ with boundary conditions $g(0)=1$, $g'(0)=0$ and check if the solution crosses the $g=0$ line and goes to $-\infty$ asymptotically. If it does then it is easy to show that there must exist at least one bound state (negative eigenvalue).
\begin{figure}[!ht]
\centering
\includegraphics[scale=0.35]{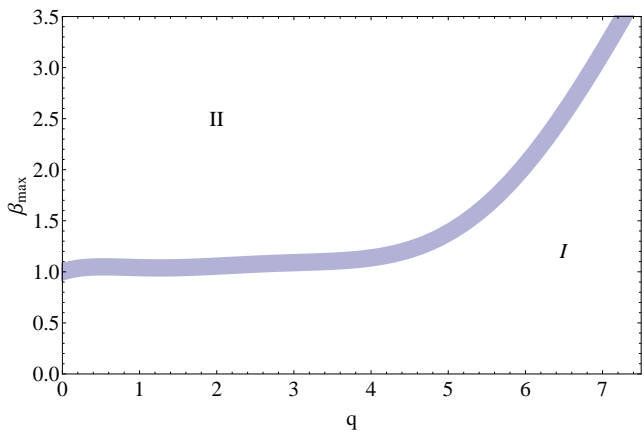}
\caption{The stability sector of the embedded dilatonic gauged vortex is shown as sector I. The parameter values of sector II correspond to instability. Notice how the stability region of $\beta$ increases as we increase the value of $q$. The thickness of the dividing line describes numerical uncertainties. }
\label{fig:stabreg}
\end{figure}
The parameter region in the $q-\beta$ space where $\hat O$ has no negative eigenvalues is shown in Fig. \ref{fig:stabreg} (sector I).  In order to construct this plot, for each value of $q>0$ we find a stability region $0<\beta < \beta_{max} (q)$. As expected $\beta_{max}(q=0)=1$. Interestingly $\beta_{max}(q>0)>1$ and the stability region increases as we increase $q$. The improvement of stability is due to the fact that the effective Schrodinger potential (\ref{ptlschr}) corresponding to the operator $\hat{O}$ becomes shallower as we increase $q$  (Fig. \ref{fig:ptlform}). Thus $\hat O$ becomes less receptive to negative eigenvalues.The new repulsive term in the Schrodinger potential (\ref{ptlschr}) (proportional to $q$) originates from the dilatonic term in the Lagrangian (\ref{dilatonicembno}). This term favors energetically a lower value for the field $\Phi$ at the origin and therefore it makes the perturbation $g$ more costly energetically at $r=0$. This leads to improved stability for the dilatonic embedded gauge vortex.

We have also investigated the dilatonic embedded global vortex in the presence of an external magnetic field. This solution is obtained by replacing the covariant derivatives in the Lagrangian (\ref{dilatoniclang}) by regular ones. In the absence of a dilatonic coupling this vortex is unstable and there is no free parameter in the Lagrangian. However, in the presence of a dilatonic coupling and a gaussian external magnetic field we have shown that the embedded global vortex gets stabilized in the region where the magnetic field is present. These results will be presented elsewhere.

\begin{figure}[!ht]
\centering
\includegraphics[scale=0.55]{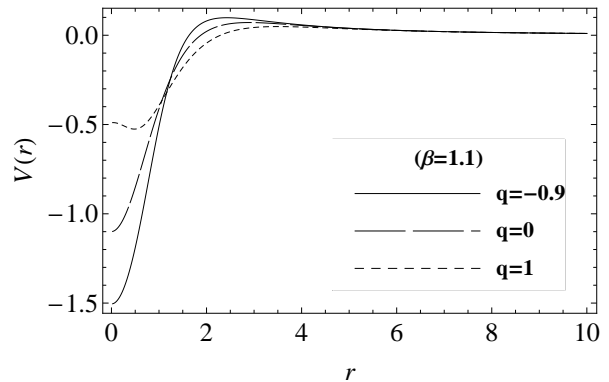}
\caption{The Schrodinger potential describing the stability of the embedded dilatonic gauged vortex. Notice that the potential becomes shallower as we increase $q$. However, it also becomes somewhat wider and this justifies the fact that for $0<q<4$ the stability improvement is very mild as shown in Fig. \ref{fig:stabreg}.}
\label{fig:ptlform}
\end{figure}

The existence of dilatonic gauge and global vortices may also have implications as a new class of models predicting spatial variation of the fine structure constant $\alpha$ on cosmological scales. Such models have been discussed in Ref. \cite{Mariano:2012wx,Mariano:2012ia,BuenoSanchez:2011wr,Olive:2012ck,Olive:2010vh} and are motivated from recent quasar absorption spectra observations that may hint towards possible spatial variation of $\alpha$ on cosmological Hubble scales \cite{King:2012id}.

The dilatonic semilocal model is perhaps the simplest generalization of the semilocal model that can lead to dramatic improvement of the semilocal vortex stability. The existence of such a dilatonic coupling in the realistic GSW model is therefore also expected to lead to improvement of the stability of the $Z$-string \cite{Achucarro:1999it,reviews} and perhaps create a stability region for the $W$-string (an alternative embedding of the NO vortex in the GSW model) \cite{Achucarro:1999it}. The analysis of the stability of the dilatonic electroweak vortices constitutes and interesting extension of the present study.

We thank Tanmay Vachaspati for useful comments and John Rizos for his help on some numerical aspects of this work. This research has
been co-financed by the European Union (European Social Fund - ESF) and Greek national funds through the
Operational Program "Education and Lifelong Learning"
of the National Strategic Reference Framework (NSRF) -
Research Funding Program: ARISTEIA. Investing in the
society of knowledge through the European Social Fund.


\section{\textbf{\emph{ Appendix}}}

In Fig. \ref{code} we show the Mathematica code used to find numerically the dilatonic vortex solutions by minimization of the energy functional (\ref{rhodilatno}). The algorithm is particularly simple and has a wide range of applications including the numerical derivation of the NO vortex solution. Note that due to stiffness of the ODE system (\ref{dilatoncnof})-(\ref{dilatonicnou}), Mathematica is unable to solve  it using the NDSolve routine. A similar code can be used to investigate the stability of the embedded dilatonic vortex by minimizing the energy functional (\ref{egdilatonic}) with respect to the three functions $f$, $u$ and $g$. For parameter values leading to a non-zero $g$ at the defect core we have instability. Even though this approach is more involved and subject to some numerical uncertainties, in most cases it leads to consistent results with the more accurate and simple perturbative method based on finding if the operator $\hat O$ has negative eigenvalues.
\begin{figure}[!ht]
\centering
\includegraphics[scale=0.65]{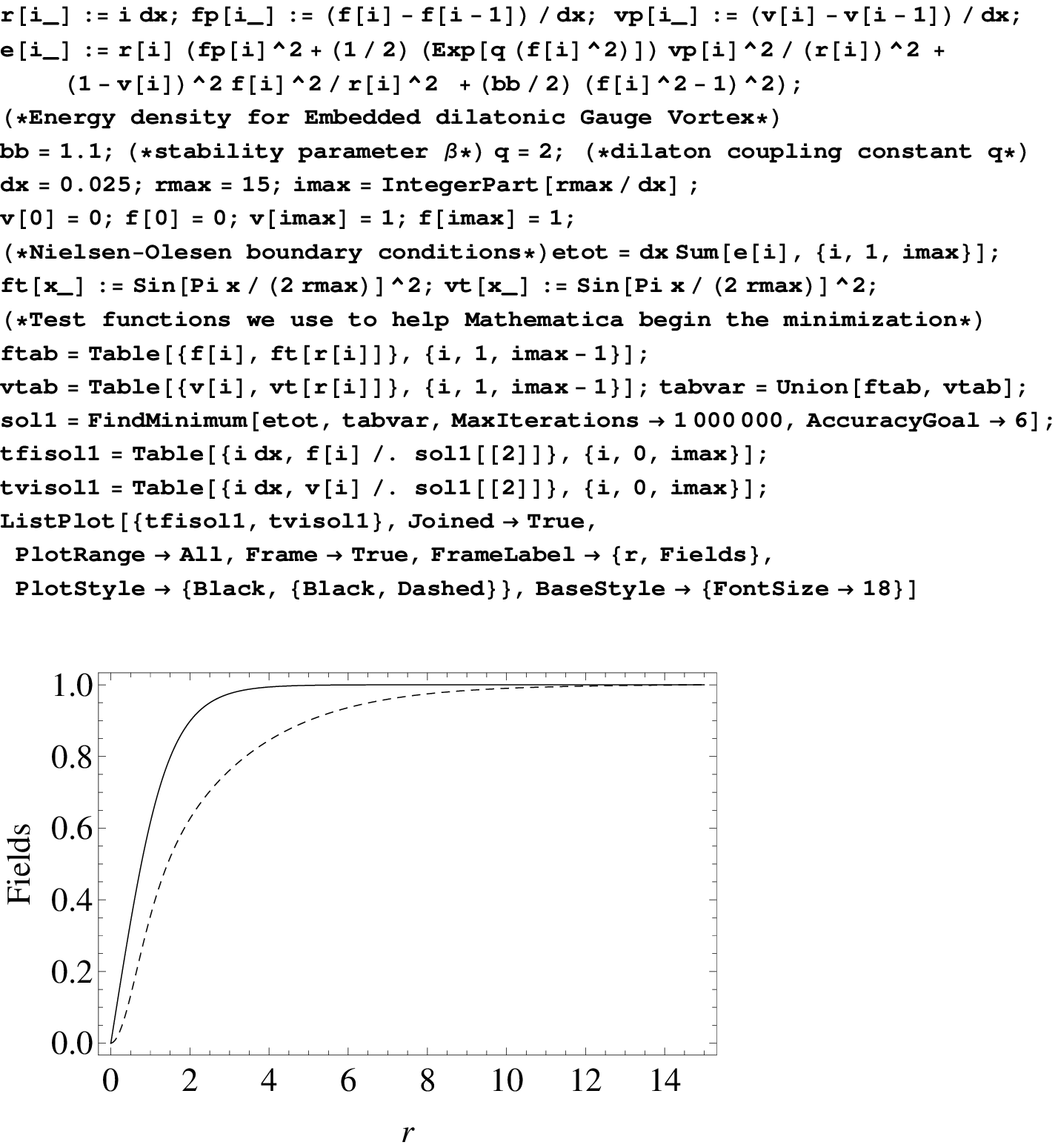}
\caption{Mathematica code for numerically finding the dilatonic vortex solutions}  \label{code}
\end{figure}


\begin{thebibliography}{999}
\bibitem{NieOle73} H. B. Nielsen and P. Olesen,
{\it Nucl. Phys.} {\bf B61} (1973) 45.

\bibitem{Peri93} L. Perivolaropoulos,
{\it Phys. Rev.} {\bf D48} (1993) 5961.

\bibitem{VacAch91} T. Vachaspati and A. Ach\'{u}carro,
{\it Phys. Rev.} {\bf D44}, 3067 (1991).

\bibitem{Hin92} M. Hindmarsh,
{\it Phys. Rev. Lett.} {\bf 68} (1992) 1263.

\bibitem{Hin93} M. Hindmarsh,
{\it Nucl. Phys.} {\bf B392} (1993) 461.

\bibitem{AchKuiPerVac92} A. Ach\'{u}carro, K. Kuijken,
L. Perivolaropoulos and T.  Vachaspati,
{\it Nucl. Phys.} {\bf B388} (1992) 435.

\bibitem{AchBorLid98} A. Ach\'ucarro, J. Borrill and A. R. Liddle,
{\it Phys. Rev. Lett.} {\bf 82} (1999) 3742.

\bibitem{GlaSalWei} G. Glashow, {\it Nucl. Phys} {\bf 22} (1961) 579;
S. Weinberg, {\it Phys. Rev. Lett.} {\bf 19} (1967) 1264; A. Salam, in
``Elementary particle physics'' (Nobel Symp. no. 8),
ed. N. Svartholm, Almqvist and Wilsell, Stockholm 1968.

\bibitem{Nam77} Y. Nambu,
{\it Nucl. Phys.} {\bf B130} (1977) 505.

\bibitem{Vachaspati:1992jk}
  T.~Vachaspati,
  Nucl.\ Phys.\ B {\bf 397}, 648 (1993).

\bibitem{JamPerVac92} M. James, L. Perivolaropoulos and T. Vachaspati,
{\it Phys. Rev.} {\bf D46} (1992) R5232;
{\it Nucl. Phys.} {\bf B395} (1993) 534.

\bibitem{HolHsuVacWat92} R. Holman, S. Hsu, T. Vachaspati
and R. Watkins,
{\it Phys. Rev.} {\bf D46} (1992) 5352.

\bibitem{EarJam93} M. A. Earnshaw and M. James,
{\it Phys. Rev.} {\bf D48} (1993) 5818.

\bibitem{VacWat93} T. Vachaspati and R. Watkins,
{\it Phys. Lett.} {\bf B318} (1993) 163.

\bibitem{GarMon95} J. Garriga and X. Montes,
{\it Phys. Rev. Lett.} {\bf 75} (1995) 2268.

\bibitem{Nac95} S. Naculich,
{\it Phys. Rev. Lett.} {\bf 75} (1995) 998.

\bibitem{EarPer94} M. A. Earnshaw and W. B. Perkins,
{\it Phys. Lett.} {\bf B328} (1994) 337.

\bibitem{Per94} L. Perivolaropoulos,
{\it Phys. Rev.} {\bf D50} (1994) 962.

\bibitem{PreVil92} J. Preskill and A. Vilenkin,
{\it Phys. Rev.} {\bf D47} (1992) 4218.

\bibitem{Achucarro:1999it}
  A.~Achucarro and T.~Vachaspati,
  Phys.\ Rept.\  {\bf 327}, 347 (2000)
  [Phys.\ Rept.\  {\bf 327}, 427 (2000)]
  [hep-ph/9904229].

\bibitem{reviews}
P. Goddard and D. Olive,
{\it Rep. Prog. Phys.} {\bf 41} (1978) 1357-1437;
T. W. B. Kibble, {\it Phys. Rep.} {\bf 67} (1980) 183;
A. Vilenkin {\it Phys. Rep.} {\bf 121} (1985) 1;
J. Preskill, ``Vortices and Monopoles'', lectures presented
at the 1985 Les Houches Summer School, Les Houches France;
A. Vilenkin and E. P. S. Shellard, ``Cosmic Strings and Other
Topological Defects'', Cambridge University Press, Cambridge (1994);
R. Brandenberger, {\it Int. J. Mod. Phys.} {\bf  A9} (1994) 2117;
M. Hindmarsh and T. W. B. Kibble, {\it Rep. Prog. Phys.}
{\bf 58} (1995) 477.

\bibitem{Mariano:2012wx}
  A.~Mariano and L.~Perivolaropoulos,
  Phys.\ Rev.\ D {\bf 86}, 083517 (2012)
  [arXiv:1206.4055 [astro-ph.CO]];

\bibitem{Mariano:2012ia}
  A.~Mariano and L.~Perivolaropoulos,
  Phys.\ Rev.\ D {\bf 87}, 043511 (2013)
  [arXiv:1211.5915 [astro-ph.CO]].

\bibitem{BuenoSanchez:2011wr}
  J.~C.~Bueno Sanchez and L.~Perivolaropoulos,
  Phys.\ Rev.\ D {\bf 84}, 123516 (2011)
  [arXiv:1110.2587 [astro-ph.CO]].

\bibitem{Olive:2012ck}
  K.~A.~Olive, M.~Peloso and A.~J.~Peterson,
  Phys.\ Rev.\ D {\bf 86}, 043501 (2012)
  [arXiv:1204.4391 [astro-ph.CO]].

\bibitem{Olive:2010vh}
  K.~A.~Olive, M.~Peloso and J.~-P.~Uzan,
  Phys.\ Rev.\ D {\bf 83}, 043509 (2011)
  [arXiv:1011.1504 [astro-ph.CO]].

\bibitem{King:2012id}
  J.~A.~King, J.~K.~Webb, M.~T.~Murphy, V.~V.~Flambaum, R.~F.~Carswell, M.~B.~Bainbridge, M.~R.~Wilczynska and F.~E.~Koch,
  MNRAS {\bf 422}, 3370 (2012)
  arXiv:1202.4758 [astro-ph.CO].

\end{thebibliography}
\end{document}